\documentclass[12pt,epsfig]{article}
\usepackage{amssymb,epsfig}

\newcommand{\half} {\frac{1}{2}}
\newcommand{\quart} {\frac{1}{4}}
\newcommand{\abs} [1] {\vert#1\vert}
\newcommand{\im} {\mathrm{Im}}
\newcommand{\sgn} {\mathrm{sgn}}

\newcommand{\Ahat} {\widehat{A}}
\newcommand{\ps} {\mathcal{P}}
\newcommand{\qs} {\mathcal{Q}}

\begin{document}

\begin{titlepage}
\hfill
\vbox{
    \halign{#\hfil         \cr
           } 
      }  
\vspace*{20mm}

\begin{center}
{\Large {\bf Large Charge Four-Dimensional\\ Non-Extremal $N=2$ Black Holes with $R^2$-Terms}\\} \vspace*{15mm}

{\sc Eyal Gruss} \footnote{e-mail: {\tt eyalgruss@gmail.com}}

\vspace*{1cm}
{\it Raymond and Beverly Sackler School of Physics and Astronomy,\\
Tel-Aviv University, Tel-Aviv 69978, Israel.\\}

\end{center}

\vspace*{8mm}

\begin{abstract}
We consider $N=2$ supergravity in four dimensions with small $R^2$ curvature corrections. We construct large charge non-extremal black hole solutions in all space, with either a supersymmetric or a non-supersymmetric extremal limit, and analyze their thermodynamic properties. This generalizes some of the extremal solutions presented in [arXiv:0902.0831]. The indexed entropy of the non-extremal extension of the supersymmetric black hole, has the form of the extremal entropy, with the charges replaced by a function of the charges, the moduli at infinity and the non-extremality parameter. This is the same behavior as in the case without $R^2$-terms.

\end{abstract}
\vskip 0.8cm

February 2009

\end{titlepage}

\setcounter{footnote}{0}

\section{Introduction}

Recently there has been an increasing interest in higher derivative $N=2$ black holes \cite{r2entropy,mohauptreview,r2stabeq,r2simple,r2interpolation,SJR,nonsusy1,nonsusy2}.
In this work we consider $N=2$ supergravity in four dimensions with small $R^2$ curvature corrections. We construct large charge non-extremal black hole solutions in all space, with either a supersymmetric or a non-supersymmetric extremal limit, and analyze their thermodynamic properties. The extremal limits were discussed in \cite{allrpaper}. The horizon geometry and entropy of small near-extremal black holes were discussed in \cite{r2ne}.

Non-extremal black holes without $R^2$-terms were discussed in \cite{nonextremal}. When considering the Bekenstein-Hawking entropy one notices that it has the same form as that of the extremal black holes, with the charges being replaced by a specific function of the charges and the non-extremality parameters. With $R^2$-terms, the entropy is no longer given by the Bekenstein-Hawking area law, but is given in general by the Wald formula \cite{wald}. A natural question to ask is whether the above property holds for the $R^2$-corrected black hole entropy. Indeed, we will provide evidence that this is the case for a class of non-extremal black holes with a supersymmetric extremal limit.

A relation between the indexed entropy of the BPS $N=2$ black holes
and the topological string partition function, evaluated at the
attractor point (horizon) has been proposed in \cite{osv}
\begin{equation}
Z_{BH} = \abs{Z_{top}}^2 \ . \label{top}
\end{equation}
We suggest that one may still use the relation (\ref{top}) for the above class of non-extremal $N=2$ black holes, with the replaced charges. This may be so at least to first order in the non-extremality parameter.
If correct, one gets all the perturbative $F$-terms corrections to the non-extremal or near-extremal $N=2$ black holes entropy using the topological string partition function. We note that this is not correct in general, as in the case of small black holes \cite{r2ne}, and in the case of black holes with a non-supersymmetric extremal limit, discussed later. Both do not exhibit the entropy structure discussed above. Note that the $R^2$-terms considered in this paper are $F$-terms. One generally expects also $D$-term corrections, which are not taken into account here. For supersymmetric black holes, it is conjectured that such terms do not contribute to the entropy \cite{osv}.

For extremal non-supersymmetric black holes, it is conjectured that the mass-charge ratio is decreased by higher curvature corrections \cite{mass_charge}. It is interesting to consider the behavior for non-extremal black holes. We will present a case where the mass-charge ratio decreases for a near-extremal black hole, and increases when the black hole is far from extremality.

We refer the reader to \cite{allrpaper} for a brief review of four-dimensional $N=2$ supergravity with $R^2$-terms, including all terminology and field definitions relevant for the current paper. A comprehensive review may be found in \cite{mohauptreview}, whose sign conventions for the curvature tensors we follow. In the paper we will use $a,b,\ldots=0,1,2,3$ to denote the tangent space
indices, corresponding to the indices $\mu,\nu,\ldots$ of the space-time coordinates $(t,r,\phi,\theta)$.

\section{Formulation of The Calculation}

We will consider $N=2$ Poincar\'e supergravity coupled to $N_V$
Abelian $N=2$ vector multiplets. The couplings of the theory are described by a prepotential, for which we assume the form:
\begin{equation}
\label{prep}
F(X,\Ahat)=\frac{D_{ABC}X^AX^BX^C}{X^0}+\epsilon\frac{D_AX^A}{X^0}\Ahat \ ,
\end{equation}
where $X^0,X^A$ are the moduli, $D_{ABC},D_A$ are constants, $D_{ABC}$ is symmetric in all indices, and $A,B,C=1\ldots N_V$. The second term, containing the scalar $\Ahat=T_{ab}^-T^{ab-}$, describes $R^2$ couplings in the Lagrangian, where $T_{ab}^-$ is an auxiliary field. This term may arise as a $g_s$ correction in the large volume limit\footnote{The large Calabi-Yau volume approximation requires $\im(X^A/X^0)\gg1$.} of type IIA string theory compactified on a Calabi-Yau three-fold, or as an $\alpha'$ correction in heterotic string theory compactified on $K3\times T^2$. We will treat the higher curvature terms in the Lagrangian as a small perturbation, in the spirit of \cite{CMP}. This is valid for the exterior region of black hole solutions in the large charge approximation. The physical dimensionless expansion parameter is one over charge squared: $Q^{-2}$. It is however convenient to express this as an expansion in $\abs{D_A}\ll\abs{D_{ABC}}$ (see also \cite{r2simple,nonsusy1,nonsusy2}). To make this explicit, we have inserted the expansion parameter $\epsilon$ in front of the second term in (\ref{prep}), and at the end we will set $\epsilon=1$. One introduces the notation: $F_I\equiv\frac{\partial}{\partial X^I}F(X^I,\Ahat),~F_{\Ahat}\equiv\frac{\partial}{\partial \Ahat}F(X^I,\Ahat)$, and similarly for higher order and mixed derivatives, where $I=0\ldots N_V$.

We look for static spherically symmetric solutions, where the metric takes the form:
\begin{equation}
ds^2=-e^{-2U(r)}f(r)dt^2+e^{2U(r)}\left(f(r)^{-1}dr^2+r^2\sin^2{\theta}d\phi^2+r^2d\theta^2\right) \ .
\end{equation}
Introduce the rescaled variables:
\begin{eqnarray}
\label{scaling}
Y^I&=&e^UX^I\nonumber\\*
\Upsilon&=&e^{2U}\Ahat\nonumber\\*
e^{-{K}}&=&i\left(\bar{Y}^IF_I(Y,\Upsilon)-Y^I\bar{F}_I(\bar{Y},\bar{\Upsilon})\right) \ .
\end{eqnarray}
The latter is called the K\"ahler potential.\footnote{The scaling $X(z)^I=e^{-{K}/2}X^I$ used in some previous works is not general enough.}

Consider black holes with one electric charge $q_0$ and $p^A$ ($A=1,2,3$) magnetic charges. In our convention $\mathcal{D}\equiv D_{ABC}p^Ap^Bp^C>0$.\footnote{The common convention uses $\mathcal{D}<0$ and a reversed sign for $q_0$.} For $q_0>0$ one can have a non-extremal solution \cite{nonextremal} with a supersymmetric extremal limit \cite{genstabeq,r2entropy,SJR}. By reversing the sign of the charge $q_0$, but taking the moduli to depend on the absolute value, one can have a non-extremal solution with a non-supersymmetric extremal limit \cite{sign_reversal,nonsusy1,nonsusy2}. At the $R$-level, i.e. without $R^2$-terms, in addition to the sign changes, the two solutions differ also in the form of the auxiliary $T$ field. The thermodynamic properties of the solution with the non-supersymmetric extremal limit can be obtained by an analytic continuation. This is no longer true when including higher curvature corrections. We will construct the $R^2$-level solutions for both cases. The extremal limits ($\mu=0$) were discussed in \cite{allrpaper}. In the following we will denote $E\equiv D_Ap^A$.

We are interested in black hole solutions of the $R^2$ curvature corrected theory, to first order in $\epsilon$. As a starting point for our ansatz, we may take the $R$-level solution ($\epsilon=0$), with the prepotential $F(\epsilon=0)$ replaced by $F(\epsilon)$. This, however, proves to be insufficient, and we need to introduce a further general linear $\epsilon$-correction to the fields. We look for solutions in the form:
\begin{eqnarray}
\label{r2sol}
e^{2U}&=&e^{-K(\epsilon)}(1+\epsilon\xi_U(r))\nonumber\\*
f&=&\left(1-\frac{\mu}{r}\right)(1+\epsilon\xi_f(r))\nonumber\\*
Y^A&=&-\frac{i}{2}y^A(1+\epsilon\xi_A(r))\qquad\textrm{(no
summation)}\nonumber\\*
Y^0&=&\half\sqrt{\frac{D_{ABC}y^Ay^By^C-4\epsilon D_Ay^A\Upsilon}{y_0}}(1+\epsilon\xi_0(r))\nonumber\\*
T_{01}^-&=&iT_{23}^-=\left(\frac{3k^3}{\alpha^3(r+k^3)}+\sgn(q_0)\frac{k_0}{\alpha_0(r+k_0)}\right)\frac{1}{r}e^{K(\epsilon)/2}(1+\epsilon\xi_T(r)) \ ,
\end{eqnarray}
where $\mu\geq0$ is a non-extremality parameter, $(k_0,k^A)>0$ are constants with either $k^1=k^2=k^3$ or otherwise only $D_{333}\neq0$, and
\begin{eqnarray}
y^A&\equiv&\frac{\alpha^Ap^A}{k^A}+\frac{\alpha^Ap^A}{r}\qquad\textrm{(no
summation)}\nonumber\\*
y_0&\equiv&\frac{\alpha_0\abs{q_0}}{k_0}+\frac{\alpha_0\abs{q_0}}{r} \ ,
\end{eqnarray}
\begin{equation}
(\alpha_0,\alpha^A)\equiv\left(\sqrt{\frac{k_0}{k_0+\mu}},\sqrt{\frac{k^A}{k^A+\mu}}\right)\qquad\textrm{(no
summation)} \ .
\label{alpha}
\end{equation}
We use
\begin{equation}
\Upsilon=-4e^{2U}(T_{01}^-)^2=-4\left(\frac{3k^3}{\alpha^3(r+k^3)}+\sgn(q_0)\frac{k_0}{\alpha_0(r+k_0)}\right)^2\frac{1}{r^2}+O(\epsilon) \ ,
\end{equation}
which is a sufficient approximation since $\Upsilon$ always comes with a factor of $\epsilon$.

The event horizon is located at $r=\mu$ and the inner horizon at $r=0$. In order for the perturbative expansion to be valid, we require $\abs{\epsilon\xi(r)}\ll1$ for all $\xi$-functions for $r\geq\mu$. We were not able to extend this requirement to $0\leq r<\mu$, as our solutions for $\xi_A(r),\xi_0(r)$ and sometimes $\xi_T(r)$ diverge at $r=0$ for $\mu>0$. However we do require that the overall $\epsilon$ correction to $g_{tt}$ satisfies a similar condition for $0\leq r<\mu$, to get the expected causal structure.\footnote{In addition, the curvature is then regular also at $r=0$.}
Although the approximation breaks down at the inner horizon, the solutions are valid in the physical region of interest, from the event horizon to infinity. In addition we set the boundary conditions: $\lim_{r\rightarrow\infty}\xi(r)=0$. This gives an asymptotically flat solution.

The equations of motion for the vector field strengths were derived in \cite{r2ne,allrpaper}. For a static spherically symmetric solution with our choice of charges, and the complex-valued form of the prepotential (\ref{prep}) and the ansatz (\ref{r2sol}) we get
\begin{eqnarray}
F_{01}^{-0}&=&iF_{23}^{-0}=\frac{1}{2F_{00}}\Big(iG_{230}-iF_{0A}F_{23}^A+\half\left(F_0+F_{0I}\bar{X}^I-64F_{\Ahat0}(2C_{0101}-D)\right)T_{01}^-\Big)\nonumber\\*
F_{01}^{-A}&=&iF_{23}^{-A}=\frac{i}{2}F_{23}^A \ ,
\end{eqnarray}
where $C_{0101}$ is the component of the Weyl tensor, $D$ is an auxiliary field and
\begin{eqnarray}
F_{23}^A&=&\frac{1}{r^2}e^{-2U}p^A\nonumber\\*
G_{230}&=&\frac{1}{r^2}e^{-2U}q_0 \ .
\end{eqnarray}

There are several other fields\footnote{Here $i,j,\alpha=1,2$, and $\Gamma=1\ldots2r$, where $r$ is an integer.} for which we consider the following truncation. We refer the reader to \cite{allrpaper} for a discussion.
\begin{equation}
A_a=\mathcal{V}_{a\phantom{i}j}^{\phantom{a}i}=Y_{ij}^I=V_a=M_{ij}=\Phi^i_{\phantom{i}\alpha}-\delta_{\alpha}^i=\mathcal{D}_aA_i^{\phantom{i}\Gamma}=\chi+2=0 \ .
\end{equation}
This implies
\begin{equation}
D=-\frac{1}{3}R \ ,
\label{DR}
\end{equation}
where $R$ is the Ricci scalar.

For our ansatz to constitute a solution, it must satisfy the
equations of motion for the metric\footnote{For a discussion on the derivation of the metric field equations, see \cite[appendix B]{r2ne}.}, the moduli
$Y^I$, the auxiliary field $T_{01}^-$, as well as the fields $A_a$, and either $V_a$ or $D$ depending on the compensating multiplet used.
In the cases that we solved, we observed that when using the hypermultiplet compensating multiplet, the equation of motion for $D$ has an overall factor of $(k^3-k_0)^2$, after substituting the ansatz. For $k^3=k_0$, one has to take this limit only after solving the equations of motion, in order not to lose a constraint.

The Einstein-Hilbert term in the Lagrangian, determines that ``Newton's constant'' is given by the unscaled K\"ahler potential:
\begin{equation}
\label{GN}
G_N^{-1}=i\left(\bar{X}^IF_I(X,\Ahat)-X^I\bar{F}_I(X,\Ahat)\right)=1-\epsilon\xi_U(r) \ .
\end{equation}
Usually one fixes $G_N=1$ as the dilatational $D$-gauge choice. This, however, is too restrictive and does not always allow a solution. Therefore $G_N$ is a function of the radial coordinate, resembling the case of dilaton gravity. The metric in the Einstein frame is given by $g^E_{\mu\nu}=G_N^{-1}g_{\mu\nu}$.
The ADM mass (in Planck units) for a non-normalized metric is given by
\begin{equation}
g^E_{tt}\big|_{r\rightarrow\infty}=g^E_{tt}(\infty)\left(1-g^E_{rr}(\infty)^{-1/2}\frac{2M}{r}+O(\frac{1}{r^2})\right) \ .
\end{equation}
One may see this by applying the coordinate transformation $t\rightarrow(-g^E_{tt}(\infty))^{-1/2}t,~r\rightarrow g^E_{rr}(\infty)^{-1/2}r$, to get the conventionally normalized line element.

The central charge is given by
\begin{equation}
Z=\lim_{r\rightarrow\infty}e^{K/2}(p^IF_I(Y,\Upsilon)-q_IY^I) \ .
\end{equation}
$Z$ is determined by the charges and the asymptotic moduli values at infinity, and does not receive higher curvature corrections.
It reads
\begin{eqnarray}
\label{Z}
\abs{Z}&=&\frac{\sqrt{2}}{4}(\abs{h_0}D_{ABC}h^Ah^Bh^C)^{1/4}\abs{3\sqrt{k^3(k^3+\mu)}+\sgn(q_0)\sqrt{k_0(k_0+\mu)}}=\nonumber\\*
&=&\quart\abs{3\sqrt{k^3(k^3+\mu)}+\sgn(q_0)\sqrt{k_0(k_0+\mu)}} \ ,
\end{eqnarray}
where
\begin{equation}
(h_0,h^A)\equiv\left(\frac{\alpha_0q_0}{k_0},\frac{\alpha^Ap^A}{k^A}\right)\qquad\textrm{(no summation)} \ ,
\end{equation}
and where in the second equality of (\ref{Z}) we imposed the normalization $g_{tt}(\infty)=-1$:
\begin{equation}
\label{norm}
4\abs{h_0}D_{ABC}h^Ah^Bh^C=1 \ .
\end{equation}
The supersymmetry algebra requires $M\geq\abs{Z}$. Note that $Z$ will not depend on $\mu$ when written as a function of the charges and $(h_0,h^A)$ alone.

The Wald entropy for the non-extremal black hole with $R^2$-terms was derived in \cite{r2ne}\footnote{This derivation uses the non-linear compensating multiplet. One may also derive the entropy formula when using the hypermultiplet compensating multiplet. This will yield the same formula, after substituting $\chi$ which is found from the equation of motion for $D$, given in \cite[eq. (4.16)]{r2stabeq}.}:
\begin{equation}
S=\lim_{r\rightarrow\mu}\left(\frac{A}{4G_N}-4A\cdot\im\Big(F_{\Ahat}(\abs{T_{01}^-}^2+16C_{0101}+16D)\Big)\right) \ ,
\label{genentropy}
\end{equation}
where $A$ is the area of the event horizon.

The Hawking temperature for a static spherically symmetric black hole is given by
\begin{equation}
T=\lim_{r\rightarrow\mu}\frac{-\partial_r g_{tt}}{4\pi\sqrt{-g_{tt}g_{rr}}} \ .
\end{equation}
One may check that after imposing (\ref{norm}), one has the first law of thermodynamics relation:
\begin{equation}
T=\frac{\partial M}{\partial S}=\frac{\partial M(q_I,p^I,h_I,h^I,\mu)}{\partial\mu}\left(\frac{\partial S(q_I,p^I,h_I,h^I,\mu)}{\partial\mu}\right)^{-1} \ ,
\end{equation}
where $q_I,p^I,h_I,h^I$ are held fixed, and assuming $\partial S/\partial\mu\neq0$.

In the following, we will further assume $k\equiv k^1=k^2=k^3=k_0$. At the $R$-level, this gives non-extremal version of the double-extremal black hole. We were able to find similar solutions for different combinations of $D_{ABC}$'s and $D_A$'s, providing that for each term such as $D_3$ there is a at least one corresponding term $D_{AB3}$. Our calculations were done using Maple with GRTensor.

\subsection{Non-Extremal Black Holes with a Supersymmetric Extremal Limit}

For simplicity, we consider the general cases with one or two indices: $D_{ABC},D_A,~(A,B,C)=1,3$; and a specific case with three indices: $D_{ABC}=D_{123},D_A=D_1,D_2,D_3$.

In the non-extremal extension of the extremal supersymmetric case ($q_0>0$), the solution reads
\begin{eqnarray}
\xi_U(r)&=&\xi_f(r)=\xi_T(r)=0\nonumber\\*
\xi_0(r)&=&\frac{E}{\mathcal{D}}\Xi(r) \ ,
\end{eqnarray}
where
\begin{eqnarray}
\Xi(r)&=&\frac{16\mu^2\left(2\ln\frac{r+k}{r}(r+k)(2k+\mu)^2(r\mu+2k r-k\mu)-k(k+\mu)P(r)\right)}{k^2(r+k)^2(k+\mu)^2(2k+\mu)}\nonumber\\*
P(r)&=&18rk^2+10rk\mu+2r\mu^2+12k^3+k^2\mu-k\mu^2 \ .
\end{eqnarray}
For $D_{ABC}=D_{333}$ and $D_A=D_3$:
\begin{equation}
\xi_3(r)=\frac{E}{3\mathcal{D}}\Xi(r) \ ,
\end{equation}
and $\xi_1(r),\xi_2(r)$ are irrelevant.
For $D_{ABC}=D_{111},D_{113},D_{133},D_{333}$ and $D_A=D_1,D_3$:
\begin{eqnarray}
\xi_1(r)&=&\frac{-D_1(D_{133}p^1+D_{333}p^3)+D_3(D_{113}p^1+D_{133}p^3)}{3((D_{113}^2-D_{111}D_{133})(p^1)^2+(D_{113}D_{133}-D_{111}D_{333})p^1p^3+(D_{133}^2-D_{113}D_{333})(p^3)^2)p^1}\Xi(r)\nonumber\\*
\xi_3(r)&=&\frac{D_1(D_{113}p^1+D_{133}p^3)-D_3(D_{111}p^1+D_{113}p^3)}{3((D_{113}^2-D_{111}D_{133})(p^1)^2+(D_{113}D_{133}-D_{111}D_{333})p^1p^3+(D_{133}^2-D_{113}D_{333})(p^3)^2)p^3}\Xi(r) \ ,\nonumber\\*
\end{eqnarray}
and $\xi_2(r)$ is irrelevant.
For $D_{ABC}=D_{123}$ and $D_A=D_1,D_2,D_3$:
\begin{eqnarray}
\xi_1(r)&=&\frac{E-2D_1p^1}{\mathcal{D}}\Xi(r)\nonumber\\*
\xi_2(r)&=&\frac{E-2D_2p^2}{\mathcal{D}}\Xi(r)\nonumber\\*
\xi_3(r)&=&\frac{E-2D_3p^3}{\mathcal{D}}\Xi(r) \ .
\end{eqnarray}
As discussed before, for $\mu>0$ our solutions for $\xi_A(r),\xi_0(r)$ diverge at $r=0$.

The entropy is given by (\ref{genentropy}), where in our solutions $D$ vanishes on the horizon to first order in $\epsilon$, and does not contribute. The entropy is
\begin{eqnarray}
S&=&2\pi\sqrt{q_0\mathcal{D}}\left(1+\frac{\mu}{k}+\frac{128E}{\mathcal{D}}\right)+O(Q^{-2})\approx\nonumber\\*
&\approx&2\pi\sqrt{\qs_0D_{ABC}\ps^A\ps^B\ps^C+256\qs_0D_A\ps^A} \ ,
\end{eqnarray}
where
\begin{equation}
(\qs_0,\ps^A)\equiv\left(\frac{q_0}{\alpha_0},\frac{p^A}{\alpha^A}\right)\qquad\textrm{(no
summation)} \ ,
\end{equation}
and $(\alpha_0,\alpha^A)$ are given by (\ref{alpha}). This has the same form as the extremal supersymmetric entropy \cite{r2entropy} with the charges $(q_0,p^A)$ replaced by $(\qs_0,\ps^A)$, which are functions of the original charges, the asymptotic moduli and the non-extremality parameter. This is the same behavior as seen at the $R$-level \cite{nonextremal}. Note also that the higher curvature correction to the entropy, does not depend on $\mu$, nor on the asymptotic moduli at infinity, similar to case of small near-extremal black holes \cite{r2ne}.
The Hawking temperature reads
\begin{equation}
T=\frac{k\mu}{8\pi\sqrt{q_0\mathcal{D}}(k+\mu)}+O(Q^{-5}) \ ,
\end{equation}
the central charge is
\begin{equation}
\abs{Z}=\sqrt{2}(q_0\mathcal{D})^{1/4}=\sqrt{k(k+\mu)} \ ,
\end{equation}
and the ADM mass takes the form:
\begin{equation}
M=\frac{\sqrt{2}(q_0\mathcal{D})^{1/4}(2k+\mu)}{2\sqrt{k(k+\mu)}}+O(Q^{-3})=k+\half\mu+O(Q^{-3}) \ .
\end{equation}
The temperature and the mass do not get first order corrections.

\subsection{Non-Extremal Black Holes with a Non-Supersymmetric\\ Extremal Limit}

In the non-extremal extension of the extremal non-supersymmetric case ($q_0<0$), the solution reads
\begin{eqnarray}
\xi_U(r)&=&\frac{64Ek^2(k+\mu)^2}{\mathcal{D}(r+k)^4}\nonumber\\*
\xi_f(r)&=&\frac{8EP_f(r)}{15\mathcal{D}k(r+k)^4(k+\mu)}\nonumber\\*
\xi_0(r)&=&\frac{4E\left(-6(r+k)^3(2k+\mu)^2(2rk+r\mu-k\mu)(5k^2+5k\mu-18\mu^2)\ln\frac{r+k}{r}+k(k+\mu)P_0(r)\right)}{15\mathcal{D}k^2(r+k)^4(k+\mu)^2(2k+\mu)}\nonumber\\*
\xi_T(r)&=&\frac{4E\left(6(r+k)^3(2k+\mu)^2(2rk+r\mu-k\mu)(5k^2+5k\mu-18\mu^2)\ln\frac{r+k}{r}+k(k+\mu)P_T(r)\right)}{15\mathcal{D}k^2(r+k)^4(k+\mu)^2(2k+\mu)} \ ,\nonumber\\*                       \end{eqnarray}
where the polynomials $P(r)$ are given in the appendix.
For $D_{ABC}=D_{333}$ and $D_A=D_3$:
\begin{equation}
\xi_3(r)=\frac{4E\left(-6(r+k)^3(2k+\mu)^2(2rk+r\mu-k\mu)(5k^2+5k\mu-18\mu^2)\ln\frac{r+k}{r}+k(k+\mu)P_{3a}(r)\right)}{45\mathcal{D}k^2(r+k)^4(k+\mu)^2(2k+\mu)} \ ,
\end{equation}
and $\xi_1(r),\xi_2(r)$ are irrelevant.
For $D_{ABC}=D_{123}$ and $D_A=D_3$:
\begin{eqnarray}
\xi_3(r)&=&\frac{4E\left(-6(r+k)^3(2k+\mu)^2(2rk+r\mu-k\mu)(35k^2+35k\mu+34\mu^2)\ln\frac{r+k}{r}+k(k+\mu)P_{3b}(r)\right)}{15\mathcal{D}k^2(r+k)^4(k+\mu)^2(2k+\mu)}\nonumber\\*
\xi_1(r)&=&\xi_2(r)=\nonumber\\*
&=&\frac{4E\left(6(r+k)^3(2k+\mu)^2(2rk+r\mu-k\mu)(15k^2+15k\mu+26\mu^2)\ln\frac{r+k}{r}+k(k+\mu)P_1(r)\right)}{15\mathcal{D}k^2(r+k)^4(k+\mu)^2(2k+\mu)} \ .\nonumber\\*
\end{eqnarray}
As discussed before, for $\mu>0$ our solutions for $\xi_A(r),\xi_0(r),\xi_T(r)$ diverge at $r=0$. Note also that for $\mu=0$ the radial derivatives of our solutions for $\xi_A(r),\xi_0(r),\xi_T(r)$ diverge at $r=0$. The curvature, however, is regular.

The entropy is given by (\ref{genentropy}), where in our solutions $D$ does not contribute to first order in $\epsilon$. Note also that $G_N(\mu)\neq1$.
The entropy reads
\begin{equation}
S=2\pi\sqrt{-q_0\mathcal{D}}\left(1+\frac{\mu}{k}+\frac{16E(75k^4+270k^3\mu+150k^2\mu^2+35k\mu^3+11\mu^4)}{15\mathcal{D}k^2(k+\mu)(2k+\mu)}\right)+O(Q^{-2}) \ ,
\end{equation}
the Hawking temperature is
\begin{equation}
T=\frac{k\mu}{8\pi\sqrt{-q_0\mathcal{D}}(k+\mu)}\left(1-\frac{8E(210k^4+30k^3\mu-40k^2\mu^2+13k\mu^3+11\mu^4)}{15\mathcal{D}k(k+\mu)^2(2k+\mu)}\right)+O(Q^{-5}) \ ,
\end{equation}
the central charge is
\begin{equation}
\abs{Z}=\frac{\sqrt{2}}{2}(-q_0\mathcal{D})^{1/4}=\half\sqrt{k(k+\mu)} \ ,
\end{equation}
and the ADM mass takes the form:
\begin{eqnarray}
M&=&\frac{\sqrt{2}(-q_0\mathcal{D})^{1/4}(2k+\mu)}{2\sqrt{k(k+\mu)}}\left(1-\frac{8E(18k^4+36k^3\mu-6k^2\mu^2-24k\mu^3-11\mu^4)}{15\mathcal{D}k(k+\mu)(2k+\mu)^2}\right)+{}\nonumber\\*
&&{}+O(Q^{-3})=(k+\half\mu)\left(1-\frac{8E(18k^4+36k^3\mu-6k^2\mu^2-24k\mu^3-11\mu^4)}{15\mathcal{D}k(k+\mu)(2k+\mu)^2}\right)+O(Q^{-3}) \ .\nonumber\\*
\end{eqnarray}
Note that when taking the limit $\mu\gg k$ in the above expressions, our approximation requires that $\abs{\frac{E\mu}{\mathcal{D}k}}\ll1$. In the following we assume $\mathcal{D}$ and $E$ have the same sign. The higher curvature corrections increase the entropy, decrease the temperature (for $\mu>0$), decrease the mass for small $\mu$ and increase the mass for large $\mu$. In the extremal limit $\mu=0$, this agrees with the mass-charge ratio conjecture of \cite{mass_charge}. The maximum temperature, where the specific heat diverges, is at lower $\mu$.

\section*{Acknowledgements}

I would like to thank Yaron Oz for his support and guidance. This work is supported in part by the Israeli Science Foundation center of excellence, by the Deutsch-Israelische Projektkooperation (DIP), by the US-Israel Binational Science Foundation (BSF), and by the German-Israeli Foundation (GIF).

\appendix
\section{Polynomials for Non-Extremal Black Holes\\ with a Non-Supersymmetric Extremal Limit}
\begin{eqnarray}
P_f(r)&=&r^3(4k^3+6k^2\mu+24k\mu^2+11\mu^3)+r^2(25k^4+44k^3\mu+96k^2\mu^2+44k\mu^3)+{}\nonumber\\*
&&{}+r(60k^5+125k^4\mu+164k^3\mu^2+66k^2\mu^3)+30k^6+60k^5\mu+65k^4\mu^2+24k^3\mu^3\nonumber\\
P_0(r)&=&r^3(236k^4+352k^3\mu-756k^2\mu^2-440k\mu^3-86\mu^4)+{}\nonumber\\*
&&{}+r^2(594k^5+771k^4\mu-2172k^3\mu^2-891k^2\mu^3-96k\mu^4)+{}\nonumber\\*
&&{}+r(448k^6+390k^5\mu-2180k^4\mu^2-510k^3\mu^3+58k^2\mu^4)+{}\nonumber\\*
&&{}+480k^7+1336k^6\mu+991k^5\mu^2+916k^4\mu^3+263k^3\mu^4\nonumber\\
P_T(r)&=&r^3(-236k^4-352k^3\mu+756k^2\mu^2+440k\mu^3+86\mu^4)+{}\nonumber\\*
&&{}+r^2(-594k^5-771k^4\mu+2172k^3\mu^2+891k^2\mu^3+96k\mu^4)+{}\nonumber\\*
&&{}+r(1472k^6+5370k^5\mu+8420k^4\mu^2+3390k^3\mu^3+422k^2\mu^4)+{}\nonumber\\*
&&{}-240k^7-1456k^6\mu-2311k^5\mu^2-2236k^4\mu^3-623k^3\mu^4\nonumber\\
P_{3a}(r)&=&r^3(216k^4+312k^3\mu-636k^2\mu^2-300k\mu^3-42\mu^4)+{}\nonumber\\*
&&{}+r^2(534k^5+651k^4\mu-1812k^3\mu^2-471k^2\mu^3+36k\mu^4)+{}\nonumber\\*
&&{}+r(388k^6+270k^5\mu-1820k^4\mu^2-90k^3\mu^3+190k^2\mu^4)+{}\nonumber\\*
&&{}+540k^7+1576k^6\mu+1471k^5\mu^2+1256k^4\mu^3+347k^3\mu^4\nonumber\\
P_{3b}(r)&=&r^3(1712k^4+2584k^3\mu+2988k^2\mu^2+1300k\mu^3+226\mu^4)+{}\nonumber\\*
&&{}+r^2(4338k^5+5757k^4\mu+6756k^3\mu^2+2583k^2\mu^3+372k\mu^4)+{}\nonumber\\*
&&{}+r(3316k^6+3090k^5\mu+3820k^4\mu^2+930k^3\mu^3+10k^2\mu^4)+{}\nonumber\\*
&&{}+300k^7-1448k^6\mu-1703k^5\mu^2-1328k^4\mu^3-331k^3\mu^4\nonumber\\
P_1(r)&=&r^3(-748k^4-1136k^3\mu-1812k^2\mu^2-800k\mu^3-134\mu^4)+{}\nonumber\\*
&&{}+r^2(-1902k^5-2553k^4\mu-4284k^3\mu^2-1527k^2\mu^3-168k\mu^4)+{}\nonumber\\*
&&{}+r(-1464k^6-1410k^5\mu-2820k^4\mu^2-510k^3\mu^3+90k^2\mu^4)+{}\nonumber\\*
&&{}+120k^7+1512k^6\mu+1587k^5\mu^2+1292k^4\mu^3+339k^3\mu^4 \ .
\end{eqnarray}

\end{document}